\documentclass[runningheads]{llncs}
\pdfoutput=1
\usepackage[T1]{fontenc}
\usepackage{textcomp}
\usepackage{graphicx}
\usepackage{float}

\usepackage{amsmath}
\usepackage{mathtools}
\usepackage{amsfonts}
\usepackage{bm} 
\usepackage{mleftright}

\newcommand{\Fitness}{\mathit{Fitness}}
\newcommand{\fmeasure}{f_{\mathit{measure}}}
\newcommand{\fdetect}{f_{\mathit{detect}}}

\newcommand{\V}[1]{\bm{#1}} 

\DeclareMathOperator*{\argmax}{arg\,max}

\DeclareMathOperator{\lsp}{lsp}
\DeclarePairedDelimiter\norm{\lVert}{\rVert}
\makeatletter
\let\oldnorm\norm
\def\norm{\@ifstar{\oldnorm}{\oldnorm*}}
\makeatother
\DeclarePairedDelimiter\abs{\lvert}{\rvert}
\makeatletter
\let\oldabs\abs
\def\abs{\@ifstar{\oldabs}{\oldabs*}}
\makeatother

\usepackage{hyperref}
\usepackage{color}

\begin{document}

\title{Adversarial Attacks on Leakage Detectors in Water Distribution Networks}
%
%
\author{Paul Stahlhofen \and
Andr\'e Artelt\orcidID{0000-0002-2426-3126} \and
Luca Hermes\orcidID{0000-0002-7568-7981} \and
Barbara Hammer\orcidID{0000-0002-0935-5591}}
\authorrunning{P. Stahlhofen, A. Artelt, L. Hermes, B. Hammer}
%
\institute{Bielefeld University, Inspiration 1, 33615 Bielefeld, Germany
\email{\{pstahlhofen,aartelt,lhermes,bhammer\}@techfak.uni-bielefeld.de}}
\maketitle              
\begin{abstract}
Many Machine Learning models are vulnerable to adversarial attacks: There
exist methodologies that add a small (imperceptible) perturbation to an input such that the model comes up with a wrong prediction. Better understanding of such attacks is crucial in
particular for models used in security-critical domains, such as monitoring of water distribution networks, in order to devise
counter-measures enhancing model robustness and trustworthiness.

We propose a taxonomy for adversarial attacks against machine learning based leakage detectors in water distribution networks. Following up on this, we focus on a particular type of attack: an adversary searching the
\textit{least sensitive point}, that is, the location in the water network where the
largest possible undetected leak could occur. Based on a mathematical
formalization of the least sensitive point problem, we use three different
algorithmic approaches to find a solution. Results are evaluated on two
benchmark water distribution networks.
\keywords{Adversarial Attacks  \and Water Distribution Networks \and Leakage
Detectors.}
\end{abstract}
\section{Introduction} 
According to the EU guidelines on Trustworthy AI, robustness is a key
requirement of Machine Learning (ML) based systems deployed in security-critical domains~\cite{trustworthy_ai}.  Judging and improving a models' robustness requires to understand its weakness so that appropiate countermeasure can be taken (see e.g. \cite{adversarial_training1}, \cite{adversarial_training2}). 
A structural weakness of many models is the existence of specifically designed inputs, so called adversarial inputs, that can cause the model to make wrong predictions with a high confidence. These adversarial inputs, first
described in~\cite{basic_adversarials}, have been used
to expose a lack of robustness for many state of the art ML models~\cite{printed_adversarials_and_bim,adversarial_tortoise,adversarial_tshirt}.

In this paper we focus on an application domain where robustness is of utmost importance: Water Distribution Networks
(WDNs).  A high amount of annual water loss in water networks across the
world~\cite{global_nrw} and an increased likelihood of droughts due to climate
change~\cite{droughts} call for an improvement and robustification of water
management. In the context of WDNs, ML based systems are used for various tasks~\cite{valerie}
e.g. detection and localization of events such as leakages~\cite{batadal,battledim} and contaminations~\cite{zhu2022review}.
\paragraph*{Related work}
In order to gain insight into vulnerabilities of a water
distribution network, the authors of~\cite{vulnerabilities_fsp} propose a methodology for building finite
state processes to find a-priori week spots independent of any monitoring system. 
The authors of~\cite{conaml} construct physics-constrained adversarial attacks for ML models operating on
cyber-physical systems. However, in their case study they only analyze the
manipulation of sensor readings for flow sensors and do not consider pressure sensors which constitute the most common type of sensor in real world WDNs.

\paragraph*{Our contributions} In this work we make the following contributions:
\begin{itemize}
\item We propose a taxonomy for adversarial attacks against leakage detectors in WDNs.
\item We investigate one particular type of attack in more detail: The search for the \textit{least sensitive point} in a WDN, which is the location in the water network where the largest possible undetected leak could occur.
We formalize this attack and propose three algorithmic approaches, which we empirically evaluate in a case study on two benchmark WDNs.
\end{itemize}
The remainder of this work is structured as follows:
In Section ~\ref{sec:foundations}, we introduce adversarial attacks in
general, as well as some important concepts of leakage detection in WDNs. Following up on this, we give a taxonomy of adversarial attacks on leakage detectors and formalize the \textit{least sensitive
point problem} in Section ~\ref{sec:adversarials_in_wdns}. Next, in
Section~\ref{sec:case_study} we empirically evaluate our proposed methods for finding the least sensitive point in a WDN. Finally, Section~\ref{sec:results} discusses the results before we conclude
with a brief summary in Section~\ref{sec:conclusion}.

\section{Foundations}
\label{sec:foundations}
\subsection{Adversarial Attacks}
The concept of adversarial attacks was first described in the context of image classification in~\cite{basic_adversarials}: There exist small and imperceptible perturbations of the original image that fool the classifier to output a completely different high-certainty classification.
The concept of an adversarial attack can be generalized (similar to~\cite{adversarial_definition}) as follows:
\begin{definition}[Adversarial Attack]
An adversarial attack is a procedure to create inputs to a Machine Learning
model, that will cause the model to make a mistake.
\end{definition}
It is important to note that the procedure for constructing an adversarial does not necessarily operate in the input space directly. In particular, consider a ML model that uses measurements as inputs:
These measurements reflect the state of a system monitored by the model.
Procedures that change the system in order to create measurements that will
cause the model to make a mistake are also considered as adversarial attacks
here. These attacks have the advantage that they are intrinsically
physics-constrained: When changing measurements directly, one has to make sure
that the resulting measurements obey physical laws, as discussed in~\cite{conaml}. Making changes to the system itself will
automatically result in measurements that reflect physical processes.

\subsection{Leakage Detection in Water Distribution Networks}
Leakage detectors typically rely on measurements of hydraulic variables such
as pressure and flow at certain points in a network. Measurements are taken at
discrete timesteps and depend on various factors, including network topology,
water consumption at network nodes and potential leakages. When creating an
adversarial input to the detector, we are mainly interested in how the
pressure values are influenced by leaks. Hence, we consider measurements based
on leakage events characterized by the area of the leak. All other factors
like water consumption and network topology are assumed to be fixed.

\begin{definition}[Measurement Function]
A measurement function maps a vector of leak areas $\V{a}_t \in \mathbb{R}^N$ for all $N$ nodes of a water distribution network at some timestep $t$ to pressure measurements at some nodes equipped with sensors. It is parametrized also by the timestep itself.
\begin{equation}
\begin{split}
\fmeasure: \mathbb{R}^N \times \{ 1, \hdots, T \} &\to \mathbb{R}^S\\
(\V{a}_t, t) &\mapsto \V{y}_t
\end{split}
\end{equation}
\end{definition}

Given the pressure measurements, we can now define the leakage detector as a
binary predictor

\begin{definition}[Leakage Detector]
A leakage detector is a function mapping pressure measurements to a binary output, indicating whether a leak was detected or not.
\begin{equation}
\begin{split}
\fdetect: \mathbb{R}^S &\to \{ 0, 1 \}\\
\V{y}_t &\mapsto \begin{cases*}
1 & \text{if a leak was detected}\\
0 & \text{otherwise}
\end{cases*}
\end{split}
\end{equation}
\end{definition}

\section{Adversarials in Water Distribution Networks}
\label{sec:adversarials_in_wdns}
\subsection{Taxonomy of Adversarials in WDNs}
\begin{figure}[h]
\centering
\includegraphics[height=\textheight,width=\textwidth,keepaspectratio=true]{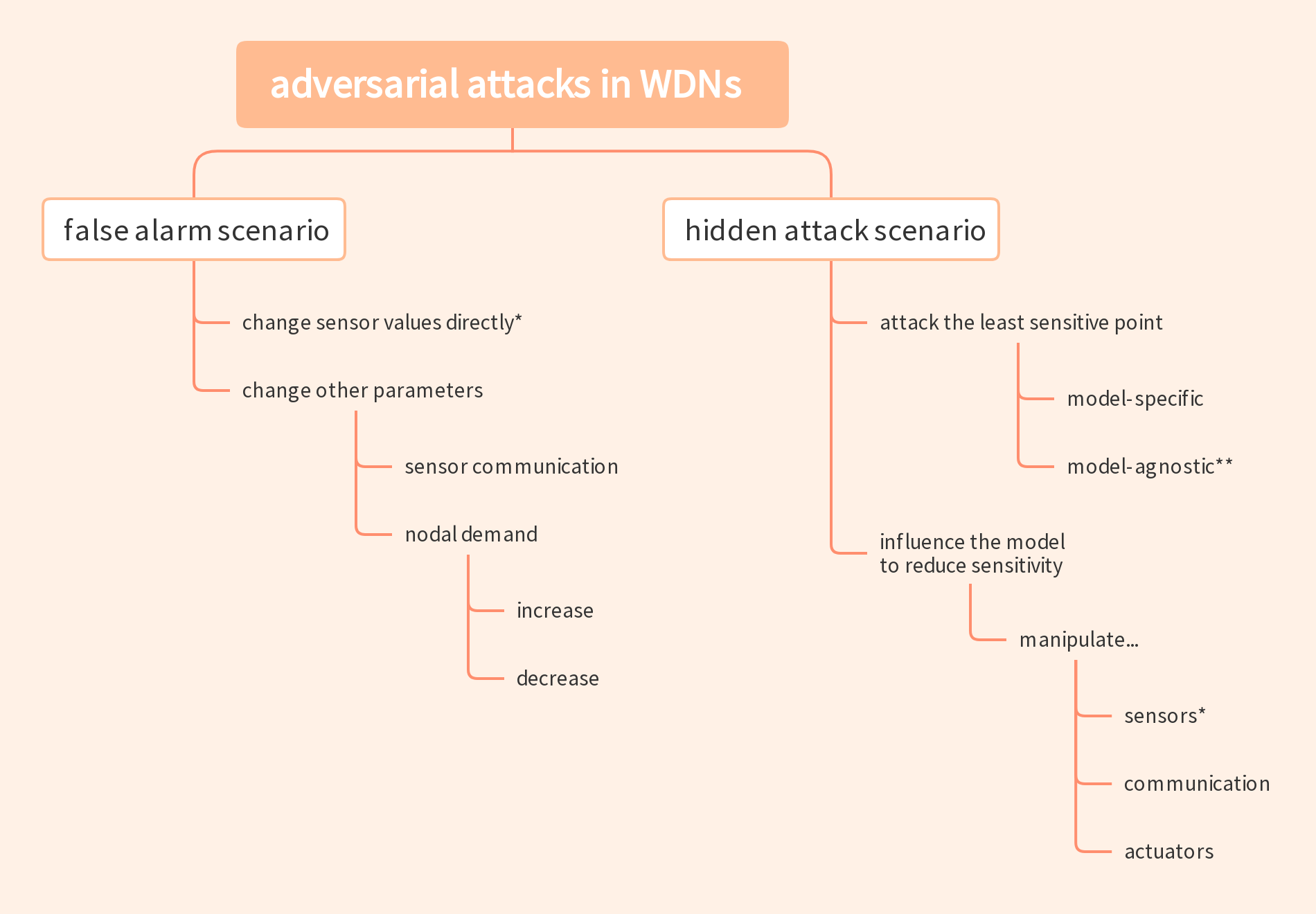}
\caption{Taxonomy showing different types of adversarial attacks in WDNs
\newline * Methods proposed in \cite{conaml}
\newline ** Metholds proposed here}
\label{fig:adversarial_taxonomy}
\end{figure}

There exist two different scenarios how a detector could be fooled:
The first one would be an input causing the
detector to predict a leak, even though no leak is actually present.  We refer
to this as the false alarm scenario;
The second one would be an input causing
the detector not to predict a leak, even though there is one.  This case can
be termed the hidden attack scenario. 

In case of the false alarm scenario, there are again two different
possibilities of attacking the system: An adversary might try to manipulate a
few network elements (e.g. sensors, pipes, etc.) they have managed to access or the attacker makes
external changes to the water demand at certain nodes.  Note that, even if the
adversary has managed to gain control over some parts of the network
infrastructure, they are not interested in actually creating a leak or other
damage in this scenario, but rather aims at causing the leakage detector to
raise a false alarm. Thus, manipulating sensor values or disturbing the
inter-sensor communication might be a promising approach. In particular, a
large decrement of pressure measurements or the complete abortion of sensory
recordings are likely to cause a wrong leakage prediction. In the attack scenario where the attacker can change node demands only, the success of the attack will depend on the number of nodes under
adversary control and the extent to which the demand can be changed at these
nodes. Generally, both a high increase in the demand e.g. by opening all water
taps in an industrial complex as well as a high decrease constitute possible
reasons for a prediction failure resulting in a false alarm.

The hidden attack scenario always involves an adversary who creates an actual
leak in the physical network. In order to fool the leakage detector
 not to find the leak, there are again two possibilities: The attacker could
try to manipulate the detector in order to make it less sensitive or they
could place the leak at a point where the detector is not very sensitive
anyway.  In the manipulation case, sensors, programmable logic controllers
(PLCs) or the sensory control and data acquisition system (SCADA) could be
targets of the attack\footnote{These constituents of a network monitoring
system were used in the Battle of the Attack Detection Algorithms (BATADAL)
\cite{batadal}. A monitoring system might have additional entry points for an
attack in practise.}. Otherwise, the adversary will have to search for points
in the network where the detector has little sensitivity to leaks.

Because investigating all the aforementioned attack scenarios in detail is behind the scope of this work, we focus on a particular type of hidden attack: Searching for locations in the network where the detector is the least sensitive to leaks.

\subsection{The Least Sensitive Point Problem}
If an adversary does not have access to sensors or other parts of the leakage detection system but still tries to create a leak that should not be detected, they have to find the location in the network where the largest undetected leak could occur. In other words, the adversary is looking for the location/point where the detector is least sensitive, which we formalize as follows:
\begin{definition}[Least Sensitive Point]\label{def:lsp}
Given a water distribution network $W = (V,E)$ with
a set of $N$ nodes $V = \{ v_n | n \in \{ 1, \hdots, N \} \}$ and
a number of $S \leq N$ pressure sensors installed in it,
a measurement function $\fmeasure$ and a leakage detector $\fdetect$,
the least sensitive point (LSP) $\lsp(W)$ in this network $W$ is defined as follows:
\begin{equation}
\begin{split}
& \lsp (W) = \argmax_{v_n \in V } \max_{\substack{t \in \{ 1, \hdots, T - K \} \\ \alpha \in \mathbb{R}^+}} \alpha\\
& \text{s.t.} \quad f_{detect} \left( f_{measure}(\alpha \V{e}_n, t+k )\right) = 0 \quad \forall k \in \{ 0, \hdots, K \}
\end{split}
\label{lsp}
\end{equation}
where $K$ is a fixed time window length and $\V{e}_n$ is the $n$-th canonical basis vector.
\end{definition}
In our experiments, we set the value of $K$ equal to the leak duration which
we fixed at 3 hours for all experiments.\\
We approach the least sensitive point problem (Definition~\ref{def:lsp}) with three algorithmic methods for two different network topologies.

\subsection{Algorithmic Approaches}
We propose the following algorithmic approaches to find the least
sensitive point (Definition~\ref{def:lsp}):
\begin{itemize}
\item Bisection Search
\item Basic Genetic Algorithm
\item Genetic Algorithm with Spectral Embeddings
\end{itemize}
We empirically evluate all three mehods in a case study in Section~\ref{sec:case_study}.

\paragraph*{Bisection Search}
The Bisection Search is based on the fact that for the least sensitive point
there has to exist some leak area $\alpha^*$, such that a leak of that area
remains undetected only at the LSP, while a leak of the same area would be
detected at every other node. Thus, one can perform a simple line search over
the leak area in order to find $\alpha^*$. Bisection search can be costly, in
particular when all timesteps in the simulation are viewed as potential
starting times of the leak. However, it yields the benefit that it is
guaranteed to find the least sensitive point in the given search space. To
reduce the computational load, we implement a pruning of nodes and starting
times based on intermediate results. Let $\alpha^i$ be the leak area after
iteration $i$. If a leak of size $\alpha^i$ remained undetected for at least
one node-time pair, we remove the following elements from the search space:
\begin{enumerate}
\item Every node for which a leak of size $\alpha^i$ was detected at every
starting time
\item Every starting time for which a leak of size $\alpha^i$ was detected at
every node
\end{enumerate}
In practise, this lead to a strong reduction of the search space after the
first iteration.

\paragraph*{Genetic Algorithm}
In the Basic Genetic Algorithm, we encode nodes and leak starting time as genes
and optimize the following fitness function
\begin{equation}
\Fitness (v_n,t) = \max \alpha \quad \text{s.t.} \fdetect(\fmeasure(\alpha
\V{e}_n, t)) = 0
\end{equation}
The largest undetected leak for each node-time pair is again computed via
a line search. In order to avoid unnecessary area maximization, we use
dynamic programming: The largest undetected leak area found so far is tested
first for every new node-time pair. If this is detected, no area maximization
was performed and the fitness is set to zero.

\paragraph*{Genetic Algorithm with Spectral Embeddings}
In order to take the network topology into account, we consider a second
version of the Genetic Algorithm, using a spectral analysis of the network's
graph Laplacian. Every node is assigned a four dimensional vector, containing
elements of the 2nd through 5th eigenvector of the graph Laplacian for that
node. Each vector element of the node embedding is then treated as a
different gene by the Genetic Algorithm. After every recombination and
mutation step, the nearest neighbour of the resulting embedding vector is
returned as offspring.\\

None of the approaches used here makes any assumptions on the type of leakage
detector. Hence, they are all model-agnostic in terms of the taxonomy (Figure
~\ref{fig:adversarial_taxonomy}).

\section{Case Study: LSP Search in Two Benchmark Networks}
\label{sec:case_study}
\subsection{Leakage Detector}
We implement a classic residual-based leakage detector~\cite{eliades2012leakage,santos2019estimation} by means of a sensor specific linear model that predicts pressure  based on the pressure measurements of all other nodes~\cite{paper_andre}:
\begin{equation}
\hat{y}_{s,t} = \V{w}_s^T \V{y}_{-s,t} + b_s \label{eq:bsi_simple}
\end{equation}
where $\V{y}_{-s,t}$ is a vector containing measurements from all sensors
except $s$ at time $t$ and $(\V{w}_s, b_s)$ are sensor specific weights
learned using data.  The residuals $r_{s,t} :=\abs{\hat{y}_{s,t}-y_{s,t}}$
determine the detector output. For the Hanoi network, we use a validation set
to learn residual weights $q_s$ for each sensor. An alram is then raised if
$\sum_s q_s r_s > 1$. In case of the L-Town network, lack of data makes the
learning of residual weights impractical. Instead, we construct thresholds for
each sensor by multiplying the maximum training error at the sensor location
with a small constant. An alarm is raised if at least one sensor residual
exceeds the corresponding threshold.

\subsection{Water Networks}
In order to obtain pressure values as input for the leakage detector, we used the hydraulic modelling software EPANET \cite{epanet} which simmulates flow and pressure values over time, given an hydraulic model of a water distribution network and water demands for each node and timestep. EPANET also allows to run simulations with leaks of arbitrary area, location, starting time and duration.

We first evaluate our proposed methodology on the Hanoi network~\cite{hanoi}. This network consists of $31$ junctions and one water reservoir from which water is entering the system through a pump.
Realistic water demands at the network nodes are generated using the
code provided with the LeakDB benchmark dataset~\cite{leakDB}.
Next, we extend our case study to the larger and more realistic L-Town network~\cite{battledim}. This network contains $782$ junctions, receiving their water from two reservoirs and one tank which is used for intermediate storage. The authors of~\cite{battledim} provided realistic demand values along with the hydraulic model of the network.

For both networks we train the leakage detector on the first five days of the timeseries and search for the least sensitive point (Definition~\ref{def:lsp}) during the two days afterwards. While detector training on realistic demands may not always be possible in advance, it is still reasonable to assume the that the first few days produced by those demands are utilized to calibrate the detector for the network at hand. In particular, infering thresholds from training errors on realistic data can help to avoid false alarms.
For the Hanoi network, we conduct a second analysis with five training days and a search space of nine days afterwards. This is not done for L-Town as simulations over a longer timeseries are computationally demanding for larger networks.

\section{Results and Discussion}
\label{sec:results}
\subsection{Hanoi Network}
For the Hanoi network, optimal solutions could be determined using the
Bisection Search. First results show that the least sensitive point is
always located close to the water source. This can be explained by the large water flow
from the reservoir to the first nodes. A leakage event along a pipe with
higher flow will lead to a less significant pressure drop when compared to a
leak along a pipe with lower flow. As this fact is known to water utility
administrators, we assume in follow up experiments that these nodes close to
the reservoir might be subject to increased protection. We treat them as
inaccessible for the adversary and remove them from the search space. 
\begin{figure}[h]
\centering
\includegraphics[height=0.3\textheight,width=\textwidth,keepaspectratio=True]{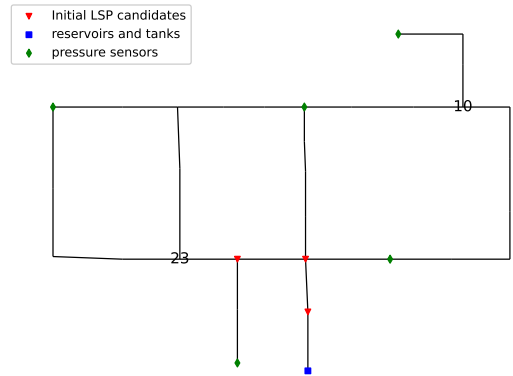}
\caption{Least Sensitive Point search on the Hanoi network: In initial
analysis, points with the lowest detector sensitivity (marked by red
triangles) are found close to the water reservoir. After excluding these
nodes from the search space, the least sensitive point is found at node 10
for the 2-days dataset and at node 23 for the 9-days dataset.
}
\label{fig:lsp_candidates_hanoi}
\end{figure}
In the 2-days dataset, the least sensitive point among the remaining nodes
is located at node 10. For the 9-days dataset, it is found to be node 23.
The results are visualized in Figure~\ref{fig:lsp_candidates_hanoi}.\\
To compare the performance between the Basic Genetic Algorithm and the extension with
spectral embeddings we run five trials of each algorithm for both datasets. In
all 20 cases, the respective algorithm is able to locate the least sensitive
point correctly. This demonstrates the high accuracy of both methods on
small water networks. In order to gain evidence for this assumption, a larger
number of trials could be conducted in future work.

\subsection{L-Town Network}
\begin{figure}[h]
\centering
\includegraphics[height=0.3\textheight,width=\textwidth,keepaspectratio=True]{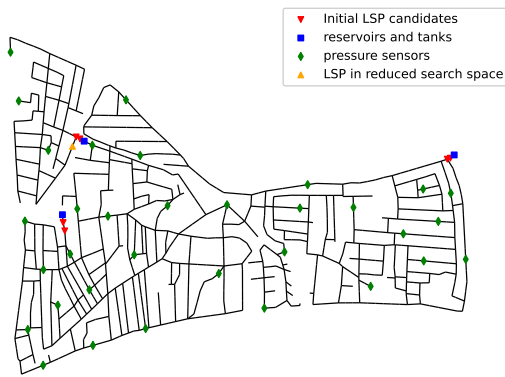}
\caption{Analysis of the L-Town network with Bisection Search for fixed
starting times. Points with a low detector sensitivity are located
close to the water sources. After excluding the initial LSP candidates from
the search space, the least sensitive point is found at node n387
(upward-pointing triangle).}
\label{fig:lsp_candidates_ltown}
\end{figure}
For the L-Town network, it is not computationaly feasible to find the least sensitive
point globally with the Bisection Search, because the network is considerably
larger than the Hanoi network and thus simulations are much more time
consuming. In order to get a realistic estimate, we determine the LSP at
three fixed leak starting times, 4, 25 and 28 hours after the beginning of the
timeseries. This starting times are assumed to have low overall demands
based on the training data, which makes leaks more difficult to detect. During the line search we are able to observe intermediate results. 
For a given leak area, we find all nodes where a leak
of that area starting at one of the fixed starting times is not detected.
Similar to the Hanoi network, we observe an agreement between those intermediate
steps for the different time trials, indicating that nodes with low
sensitivity are located close to the water sources. Figure
~\ref{fig:lsp_candidates_ltown} shows all nodes for which a leak with an
area of 100 $\text{cm}^2$ is not detected in at least one of the start time
trials as downward-pointing triangles. From the result we can also obseve that the direction of water flow plays a major role in leakage detection: Even though there is a pressure sensor
located directly next to the tank in the upper left area of the network, nodes
on the other side of the tank are very vulnerable to large undetected leaks.
This is because the sensor only measures the pressure of water flowing into
the tank and does not support the detection of downstream leaks. In case of
the L-Town network, pressure-based detection close to the water sources is
also hampered by pressure reducing valves, located after each source. The
valves smooth out incoming pressure values, making the detection of downstream
leaks more difficult.

\begin{figure}[h]
\centering
\includegraphics[width=\textwidth,height=0.3\textheight,keepaspectratio=true]{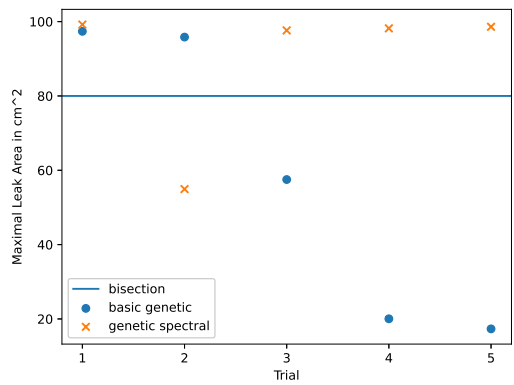}
\caption{Comparison of the three algorithmic approaches. }
\label{fig:luca_ftw}
\end{figure}

In order to compare the performance of our algorithmic approaches on the L-Town
network, we first re-compute the least sensitive point for the same starting
times used above after excluding the initial LSP candidates marked in Figure
~\ref{fig:lsp_candidates_ltown} from the search space. All three runs of
the Bisection Search agreed in node n387 as the new least sensitive point
(upward-pointing triangle in Figure ~\ref{fig:lsp_candidates_ltown}). We then
conduct five trials for each genetic algorithm.
Results are shown in Figure ~\ref{fig:luca_ftw}. The horizontal line
inidicates that each run of the Bisection Search found the least sensitive point at a
maximum undetected leak area of 80 $\text{cm}^2$. All genetic-algorithm
results with a larger leak area identify n387 as the LSP, while all results
with a smaller leak area fail to do so\footnote{In theory, it is possible that trials with a smaller leak area  still find the same node, but this is not the case here.}. The comparison suggests that the genetic algorithm with spectral node embeddings performs well at finding the LSP, also
for larger networks. A simple enhancement to reduce the uncertainty of a
single trial would be to conduct multiple successive trials, keeping the
result of the best one as global solution. This would still be computationally much
cheaper than an exhaustive Bisection Search. In future
work, a larger number of trials may be used to determine the junction accuracy
of both algorithms.

\section{Conclusion}
\label{sec:conclusion}
In this work, we applied the concept of adversarial attacks to leakage
detection in water distribution networks. For this purpose, we proposed a
taxonomy of potential adversarials against a leakage detector. We then focused
on a particular type of adversarial: Attacking the network at the least
sensitive point. We formalized the least sensitive point
problem and proposed three algorithmic approaches to solve it.
In practice, knowing the least sensitive point and the vulnerability to attacks constitutes crucial information which enables practitioners to develop more robust methods using e.g. adversarial training~\cite{bai2021recent} or an improvement of the robustness by introducing targeted sensors.

We empirically evaluated our proposed methods in a case study on two benchmark WDNs.
For the genetic algorithm with spectral embeddings, experiments yielded
promising results on both networks. This method allows much faster solutions
when compared to the Bisection Search, due to a lower runtime complexity. The
locations of the least sensitive point indicate that the detectors weak spots
are highly dependent on network topology, in particular on the location of the
water sources.

In future work, we plan to address the same problem with another leakage
detection model to compare results between different leakage detectors.
In order to increase the robustness, we will determine the
effect of incremental sensor placement on the maximum undetected leak area.
For this purpose, we will simulate sensors at the location where the least
sensitive point has previously been detected and re-run the experiments with
the new sensors in place. In this way, knowledge about adversarials can be 
helpful in practical applications.

\paragraph{Acknowledgements}
We gratefully acknowledge funding from the European Research Council (ERC)
under the ERC Synergy Grant Water-Futures (Grant agreement No. 951424).
%
%
\bibliographystyle{splncs04}
\bibliography{adversarials_vs_detectors.bib}
\end{document}